%% file: MusicJudge.tex
\documentclass[cameraready]{Interspeech}

\usepackage[table,dvipsnames]{xcolor}
\usepackage{booktabs}
\usepackage{makecell}
\usepackage[normalem]{ulem}
\usepackage{subcaption}
\usepackage{multirow}

\usepackage{pgfplots}
\pgfplotsset{compat=1.18}
\usepgfplotslibrary{groupplots}

\usepackage{fontspec}
\newfontfamily\devanagarifont[
    Path=fonts/,
    Extension=.ttf,
    Script=Devanagari
]{NotoSansDevanagari-Regular}

\usepackage{pifont}

\input{content/ours_names}

\title{Listening Like a Judge: A Music-Aware Framework for Automatic Singing Performance Evaluation}

\input{content/authors}


\usepackage{comment}

\begin{document}

\maketitle

\begin{abstract}
Automatic singing quality assessment (SQA) requires evaluating lyrical correctness and musical fidelity while handling expressive variations. However, existing systems largely rely on either acoustic cues or lyric transcriptions exclusively, limiting holistic performance evaluation. Furthermore, their integration is non-trivial due to challenges in robust singing transcription amid melisma, vibrato, and tempo elasticity. To this end, we propose {\ourmethod}, a modality-guided framework for automated SQA that performs block-aligned multimodal analysis by coupling lyric correctness with pitch–rhythm fidelity. It detects semantically meaningful lyric blocks using multi-signal matching that integrates semantic embeddings, lexical similarity, and phonetic alignment. To improve singing audio transcription, we introduce Modality-Guided LoRA for ASR fine-tuning. Experiments across datasets demonstrate strong agreement with human expert judgments and validate the generalizability of {\ourmethod}.
\end{abstract}

\vspace{-0.8em}
\section{Introduction}
\vspace{-0.2em}

Singing quality assessment (SQA) is a multifaceted problem involving lyrical accuracy, pitch intonation and rhythmic timing. Human experts evaluate vocal performances based on correct lyric pronunciation and adherence to the underlying melodic and rhythmic structure of the music (e.g., \textit{Raag} in Indian classical music). However, objective evaluation is challenging because singers often introduce acceptable variations, including pronunciation changes and deliberate creative improvisations, further compounded by singer-to-singer differences in vocal timbre, pitch range, and expressive style.
Existing computational SQA use isolated metrics like pitch deviation or lyric transcription accuracy (via automatic speech recognition or ASR), which fail to capture the holistic judgments by human evaluators. Moreover, signal-level similarity measures penalize musically valid improvisations, while text-based lyric matching overlooks phonetic and ordering variations in sung content.
In this work, we propose a block-aligned multimodal framework for automated SQA
that is resilient to stage performance nuances like audience noise, \textit{in media res}, bridge entry, etc. We analyze performances at semantically meaningful temporal segments (e.g., verses, choruses) in two complementary dimensions: \textit{content fidelity} and \textit{musical quality}. These signals are aggregated to produce an interpretable unified singing performance score. Unlike rigid pitch-threshold or transcript-matching systems, our {\ourmethod} models acceptable expressive variations while preserving musical structure. Our main contributions are:\looseness=-1
\begin{itemize}
  \vspace{-0.1em}
    \item We present the first block-aligned multimodal SQA framework that jointly models lyrical grounding and music-aware pitch–rhythm fidelity, producing interpretable scores with strong human expert correlation.
    \item We introduce a multi-signal lyric alignment and scoring mechanism that integrates semantic embeddings, fuzzy lexical matching, and phonetic similarity, allowing robust detection and evaluation of sung lyric segments even under ASR errors, pronunciation variation, and melismatic singing.
    \item We introduce Modality-Guided LoRA (MG-LoRA), a music-aware fine-tuning strategy for ASR that integrates pitch, timing, and alignment cues, significantly improving lyric transcription robustness.
\end{itemize}

\noindent\textbf{Related Work:} Early SQA methods rely on handcrafted acoustic features and shallow models~\cite{8659545}, with later neural approaches introducing temporal modeling for vocal dynamics~\cite{leng2021mbnet,10691132}. However, these methods remain \textbf{largely limited to acoustic analysis} and do not integrate musical structure with lyrical content. Musical representation learning methods capture pitch, tonality, and rhythm through structured audio embeddings and harmonic context~\cite{yuan2023marble,kang2023hclas,hsieh2025tonality}, but they \textbf{do not address lyric-aligned transcription or melismatic tokenization} in singing ASR. Recent singing transcription approaches adapt transformer-based ASR and benchmark robustness under musical variability~\cite{wu2024songtrans,tang2025singmos}, yet they \textbf{do not explicitly model pitch continuity or onset cues} to handle melisma-induced segmentation errors. More recent work leverages self-supervised audio representations for singing assessment~\cite{narang2024automatic}, but \textbf{acoustic modeling and lyric decoding remain largely decoupled}.
\newline\indent In contrast, our approach integrates pitch contour, duration stability, and onset alignment into ASR fine-tuning objective, enabling segmentation-aware transcription aligned with musical structure and linguistic decoding. To the best of our knowledge, this is the first work to jointly model these aspects for SQA. Experiments on our {\ourdata} dataset demonstrate strong agreement with human expert judgments (\textbf{Spearman correlation of 0.683, 32\%$\uparrow$}, \textbf{Kendall's $\mathbf{\tau}$ of 0.499, 41\%$\uparrow$}), while results on Jamendo ~\cite{durand-2023-contrastive} and SingMOS-Pro further demonstrate the generalizability of {\ourmethod}.

\begin{figure}[t]
    \centering
    \includegraphics[width=\linewidth, trim = 0.85cm 0cm 0.06cm 0cm, clip]{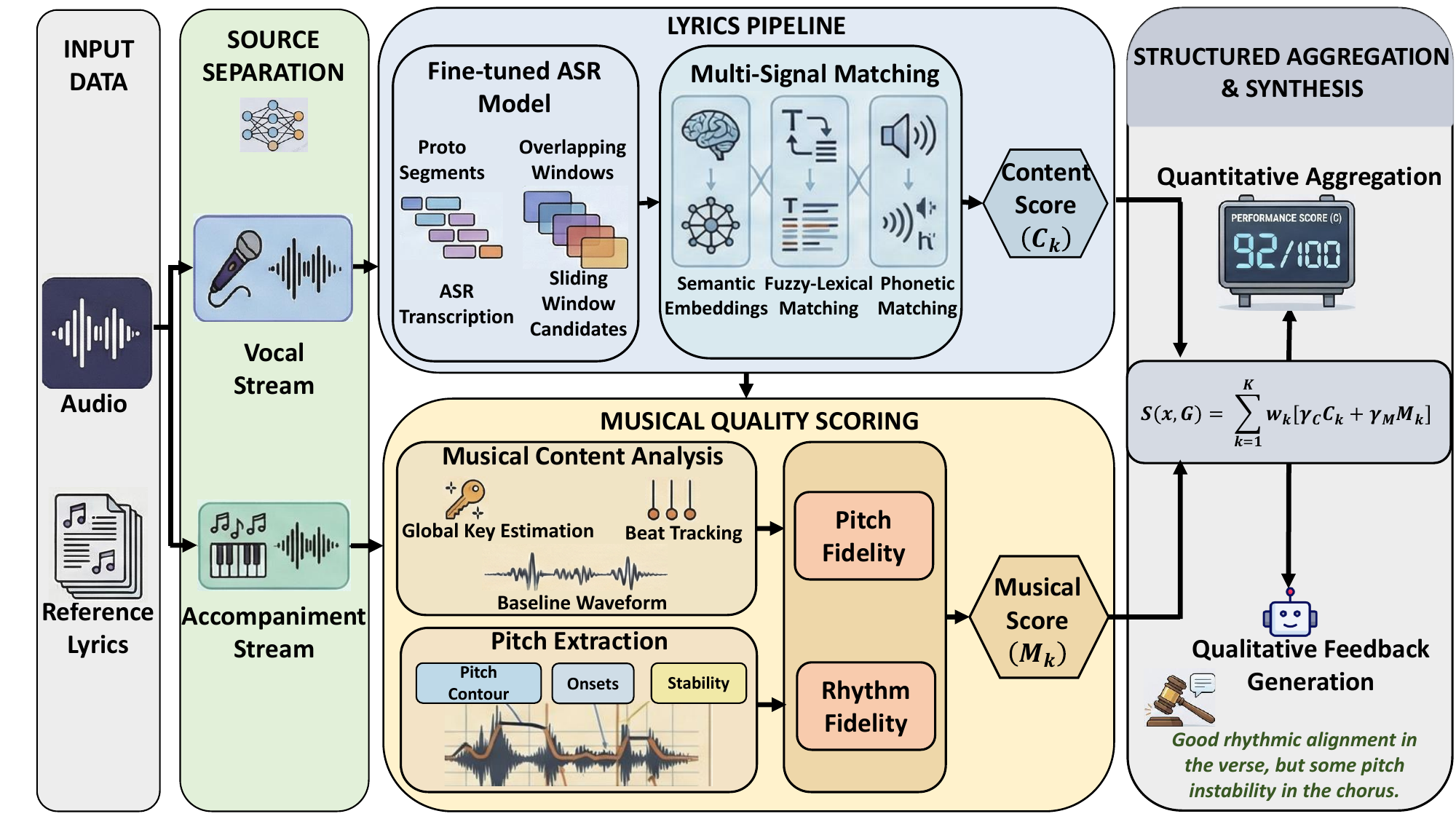}
    \vspace{-2em}
    \caption{{\ourmethod} jointly evaluates the content and musical aspects of singing audio.
    \vspace{-2em}
    }
    \label{fig:architecture}
\end{figure}

\vspace{-1em}
\section{Problem Formulation}\label{sec:problem_formulation}
\vspace{-0.3em}

Let $x(t)$ denote a singing performance waveform defined over 
$t \in [0,T]$, where $T$ is the total duration. 
Let $\mathcal{G}=\{\ell^{*}(t), \mathcal{Z}^{*}\}$ denote the global reference 
comprising ground-truth lyrics and canonical musical structure 
(e.g., tonal framework).
Source separation yields vocal and accompaniment streams:
$x(t) \rightarrow (x_v(t), x_a(t))$.

\vspace{-0.6em}
\subsection{Segmentation Under Structural Uncertainty}
\vspace{-0.1em}

\oursubsubsection{ASR-Derived Proto-Segments}
\vspace{-0.3em}
Let $\{\tilde{S}_n\}_{n=1}^{N}$ denote temporal proto-segments obtained 
from transcription of $x_v(t)$:
$\tilde{S}_n = \{ t \mid \tilde{t}_n^{(s)} \le t \le \tilde{t}_n^{(e)} \}$.
Here $\tilde{t}_n^{(s)}$ and $\tilde{t}_n^{(e)}$ are start and end times 
induced by ASR token or lyric-line boundaries (proto-segment boundaries). 
Due to singing-specific phenomena (e.g., vowel elongation, melisma, vibrato), 
these boundaries may not align with musically coherent units.

\vspace{-0.3em}
\oursubsubsection{Sliding-Window Block Candidates}
\vspace{-0.4em}
To mitigate segmentation uncertainty, overlapping candidates are formed:
$W_m = \bigcup_{n=m}^{m+L-1} \tilde{S}_n$,
$m=1,\dots,N-L+1$,
where $L$ is the window length (in proto-segments).

\vspace{-0.3em}
\oursubsubsection{Block Selection} \vspace{-0.4em}
Final evaluation blocks $\mathcal{B}=\{B_k\}_{k=1}^{K}$ are selected 
from $\{W_m\}$ based on multi-signal structural coherence, i.e., 
$B_k \in \{W_m\}$.
Each selected $B_k$ corresponds to a linguistically and musically coherent unit 
(e.g., verse, chorus, bridge, or \textit{alaap}) and admits a temporal representation,
$B_k = \{ t \mid t_k^{(s)} \le t \le t_k^{(e)} \}$,
where $t_k^{(s)}$ and $t_k^{(e)}$ denote the inferred start-end 
times obtained from the selected window $W_m$.
Subsequent evaluation operates on $B_k$.

\vspace{-0.6em}
\subsection{Block-Level Fidelity Measures}\label{sec:block_level_fidelity_measures}
\vspace{-0.2em}

For each block $B_k$, we extract --
(a) transcribed lyrics $\hat{\ell}_k$ from $x_v(t)$,
(b) pitch contour $\mathbf{p}_k(t)$ and vocal onsets $\mathbf{o}_k$ from $x_v(t)$,
(c) beat sequence $\mathbf{b}_k$ from $x_a(t)$, and
(d) global key $\mathcal{K}$ estimated once from $x_a(t)$ and shared across blocks.
Let $\ell_k^{*}$ denote reference lyrics aligned to $B_k$.

\textbf{Content Fidelity},
$\mathcal{C}_k = \sum_i \alpha_i \, s_i(\ell_k^{*}, \hat{\ell}_k)$,
$\quad \sum_i \alpha_i = 1$,
where $s_i(\cdot)$ denote complementary semantic, lexical, and phonetic 
similarity measures and $\alpha_i$ denote weighting coefficients.
\textbf{Pitch Fidelity},
$\mathcal{P}_k = 1 - \frac{1}{|B_k|}
\int_{B_k} \rho_p(\delta_p(t; \mathcal{K})) \, dt$,
where $\delta_p(t; \mathcal{K})$ denotes deviation relative to the
performance-intrinsic global key $\mathcal{K}$,
and $\rho_p(\cdot)$ is a bounded expressive penalty function.
The specific construction of $\delta_p(\cdot)$ is described in Sec.~\ref{sec:methodology}.
\textbf{Rhythmic Fidelity},
$\mathcal{R}_k =
1 - \frac{1}{|\mathbf{o}_k|}
\sum_{o_i \in \mathbf{o}_k}
\rho_r(\delta_r(o_i))$,
where $\delta_r(o_i)$ denotes normalized deviation between a vocal onset 
$o_i$ and its nearest beat in $\mathbf{b}_k$.
The \textbf{musical score} for block $B_k$ is:
$
\mathcal{M}_k = \beta_1 \mathcal{P}_k + \beta_2 \mathcal{R}_k,
\quad \beta_1 + \beta_2 = 1
$.

\vspace{-0.4em}
\subsection{Structured Aggregation}
\vspace{-0.2em}

Let $|B_k| = t_k^{(e)} - t_k^{(s)}$ denote the duration of block $B_k$.
We define duration weights:
$w_k = \frac{|B_k|}{\sum_{j=1}^{K} |B_j|},
\hfill\sum_{k=1}^{K} w_k = 1$.
The overall performance score is then given by:\vspace{-0.5em}
\begin{equation}
\mathcal{S}(x,\mathcal{G}) =
\sum\nolimits_{k=1}^{K}
w_k
\left[
\gamma_{{}_\mathcal{C}} \mathcal{C}_k +
\gamma_{{}_\mathcal{M}} \mathcal{M}_k
\right]
\label{eq:overall_performance_score}
\vspace{-1em}
\end{equation}

\vspace{0.1em}
\oursubsubsection{Probabilistic Interpretation}
Each penalty $\rho(\delta)$ corresponds to a negative log-likelihood 
under an implicit expressive noise model. 
Under conditional independence across blocks and modalities, 
$\mathcal{S}(x,\mathcal{G})$ is proportional to the log-likelihood 
of the observed performance given the reference structure, 
with block boundaries treated as latent structural variables 
inferred through multi-modal consistency.

\vspace{-0.8em}
\section{Methodology}\label{sec:methodology}
\vspace{-0.2em}

Our framework targets SQA through block-aligned 
multi-modal analysis, integrating lyrics-aware content scoring and 
pitch–rhythm modeling as depicted in Fig.~\ref{fig:architecture}.

\vspace{-0.6em}
\subsection{Dataset Curation and ASR Adaptation}\label{sec:dataset_curation_and_asr_adaptation}
\vspace{-0.2em}

To improve lyric transcription robustness, we fine-tune 
\texttt{whisper-large-v3} on singing data.
Many existing datasets \cite{tang2025singmos, stoller2019end}
lack pitch-related information,
while some have partial lyrics content ($\sim4.3k$ English lines in \cite{tang2025singmos}).
There is also a significant data gap in the coverage of musical performance
recordings alongside their reference lyrics, which is further constrained
due to copyright restrictions.
So, for this work, we also curate {\ourdata}, a corpus of 420 samples (train/val/test: 70/15/15), comprising
(a) singing performances (including audience noise, judge commentary),
(b) authoritative playback audio, and
(c) native-script lyrics.
Here, (b) and (c) serve as ground-truth references 
during evaluation. Portion used for fine-tuning is
either locally recorded by 
the authors' institutional band or appropriately licensed. 
To improve robustness to acoustic variability, we apply data 
augmentation (noise mixing, tempo perturbation).
{\ourdata} primarily consists of Indian music, particularly 
solo songs, spanning diverse moods, genres, eras, and singer demographics.

\vspace{-0.6em}
\subsection{Source Separation}
\vspace{-0.2em}

Given an input performance audio, we do source 
separation using \texttt{Demucs} \cite{rouard2023hybrid} to obtain vocal and accompaniment.
Vocal stream supports lyric-pitch analysis, while 
accompaniment stream supports beat-tonal estimation.

\vspace{-0.6em}
\subsection{Lyrics Pipeline: Reference-Guided Block Detection and Scoring}
\vspace{-0.2em}

The lyrics evaluation pipeline follows a reference-guided but 
ASR-driven progression. Block-wise analysis accommodates live performances that may begin from arbitrary song sections or reorder structural parts such as intro, verse, and chorus. We select Whisper \cite{radford2023robust}
as our base ASR model to leverage its inherent
pause-based segmentation, which is likely to yield segments parallel
to musical phrases. Each segment comprises a raw transcript with a timestamp.

\vspace{-0.3em}
\oursubsubsection{Modality-Guided Fine-Tuning for Singing ASR} \vspace{-0.4em}
To improve temporally stable and lyrics-faithful transcription under singing-specific acoustic variations,
we fine-tune \texttt{whisper-large-v3} 
on curated music data (Sec.~\ref{sec:dataset_curation_and_asr_adaptation}), leveraging Low-Rank Adaptation (LoRA).
With MG-LoRA, model is optimized using a composite objective that combines the standard sequence-to-sequence cross-entropy loss, $\mathcal{L}_{\text{ASR}}$, with authoritative lyrics as targets, augmented by singing-aware regularization terms.
Specifically, we optimize
\vspace{-0.6em}
\begin{equation}
    \mathcal{L}_{\text{total}}
    =
    \mathcal{L}_{\text{ASR}}
    +
    \lambda_d \mathcal{L}_{d}
    +
    \lambda_p \mathcal{L}_{p}
    +
    \lambda_a \mathcal{L}_{a}
    +
    \lambda_o \mathcal{L}_{o}
    \label{eq:whisper_ft_loss_fn}
\end{equation}

Here, 
$\mathcal{L}_{d}$ penalizes unstable token duration in sustained segments,
$\mathcal{L}_{p}$ discourages token boundary proliferation within acoustically smooth fundamental frequency regions,
$\mathcal{L}_{a}$ enforces monotonic alignment consistency,
and
$\mathcal{L}_{o}$ encourages token boundaries to coincide with detected vocal onset structure.
The coefficients are selected by tuning on a small validation set. For our experiments, these are $\lambda_d = 0.10$, $\lambda_p = 0.15$, $\lambda_a = 0.10$, and $\lambda_o = 0.05$.

Augmentation strategies 
(e.g., additive noise, tempo perturbation) are applied 
to improve robustness to performance variability and 
background interference (e.g., audience reactions).
We then convert it to Faster-Whisper for lower inference latency.
The separated input vocal stream is then transcribed using the 
fine-tuned Whisper model. The output consists of time-aligned 
tokens grouped into proto-segments. Since no explicit structural 
labels (e.g., verse, chorus) are available at inference time, 
song structure is treated as latent.

\vspace{-0.4em}
\oursubsubsection{Sliding ASR Window Creation} \vspace{-0.4em}
Boundary distortions caused by singing-specific acoustic effects 
(e.g., vowel elongation, melisma, vibrato) may misalign 
ASR token boundaries from musically coherent units.
Furthermore, Whisper operates in 30s windows, further
introducing undesired boundaries.
To mitigate this, we group proto-segments into overlapping sliding windows 
to generate candidate temporal text blocks.

\vspace{-0.3em}
\oursubsubsection{Multi-Signal Block Detection}\label{sec:multi_signal_block_detection} \vspace{-0.4em}
Each candidate window is compared against reference lyrics 
using complementary similarity signals:
(a) \textbf{embedding similarity:} sentence-level 
semantic embeddings measure contextual alignment.
(b) \textbf{fuzzy lexical matching:} normalized 
edit-distance captures surface-form correctness.
(c) \textbf{phonetic matching:} grapheme-to-phoneme 
conversion enables pronunciation-aware alignment.
Windows are assigned to reference blocks based on joint 
multi-signal coherence, thereby refining block boundaries 
through reference-guided alignment rather than fixed segmentation.

\vspace{-0.4em}
\oursubsubsection{Line-Level Ordered Matching} \vspace{-0.4em}
Within each detected block, line-level sequential alignment 
is done to ensure correct progression. We employ HIT/MISS 
to detect missing lines, repeated/spurious lines, and ordering inconsistencies.
This captures structural coverage and lyrical flow 
beyond surface similarity.

\vspace{-0.4em}
\oursubsubsection{Block Content Scoring} \vspace{-0.4em}
For each block, three normalized measures are computed -- (a) \textbf{coverage:} proportion of reference lines correctly detected, (b) \textbf{correctness:} lexical and phonetic fidelity, and (c) \textbf{flow:} sequential consistency and order preservation.
These are combined to produce a block-level content score, $\mathcal{C}_k$, 
which contributes to the overall lyrics score after aggregation 
across blocks.

\vspace{-0.6em}
\subsection{Musical Quality Scoring}
\label{subsec:pitch_rhythm_method}
\vspace{-0.2em}

We design {\ourmethod} to reward adherence to melodic principles
without penalizing creative deviations. Notably, we do not
constrain the performance to be grounded to $\mathcal{Z}^{*}$.
Next, we describe the pitch and rhythm deviation terms
$\delta_p(\cdot)$ and $\delta_r(\cdot)$ introduced in Sec.~\ref{sec:problem_formulation}.

\vspace{-0.4em}
\oursubsubsection{Global Key Estimation} \vspace{-0.4em}

A single global key $\mathcal{K}$ is estimated from the accompaniment
signal $x_a(t)$ using chroma-based tonal profile matching.
$\mathcal{K}$ is inferred once per performance and shared
across all blocks $\{B_k\}$.
This is done from the accompaniment of the
\emph{input performance} rather than the playback reference $\mathcal{Z}^{*}$ to
avoid penalizing intentional transposition while still enforcing
intra-performance tonal consistency.

\vspace{-0.4em}
\oursubsubsection{Pitch Deviation Construction} \vspace{-0.4em}

For each block $B_k$, the vocal pitch contour $\mathbf{p}_k(t)$
is extracted from $x_v(t)$ using Probabilistic YIN (pYIN) \cite{6853678}.
Voiced frames are retained via pYIN masking.
We compute three complementary components --
(a) \textbf{in-key deviation:} minimum circular distance between pitch class $c(t)$ and the scale induced by $\mathcal{K}$,
(b) \textbf{stability:} short-term variance within sustained regions,
and, (c) \textbf{voiced rate:} proportion of voiced frames within $B_k$.

The aggregated pitch deviation for block $B_k$ is defined as:
$
\delta_p^{(k)} =
\lambda_1 \overline{d}_{\text{scale}}^{(k)}
+
\lambda_2 \overline{\sigma}^{(k)}
+
\lambda_3 (1 - v_k),
\quad \sum_i \lambda_i = 1
$,
where $\overline{d}_{\text{scale}}^{(k)}$, $\overline{\sigma}^{(k)}$,
and $v_k$
denote block-averaged in-key distance, stability,
and voiced-frame ratio, respectively.
The pitch fidelity score is obtained via bounded normalization:
$\mathcal{P}_k = 1 - \rho_p\left(\delta_p^{(k)}\right)$,
where $\rho_p(\cdot)$ applies clipping-based normalization to
ensure $\mathcal{P}_k \in [0,1]$.

\vspace{-0.4em}
\oursubsubsection{Rhythmic Deviation Construction} \vspace{-0.4em}

For each onset $o_i \in \mathbf{o}_k$, we compute normalized
beat-alignment deviation:
$\delta_r(o_i) = \frac{|o_i - \text{NN}(o_i; \mathbf{b}_k)|}{\tau_k}$,
where $\tau_k$ is the local inter-beat interval.
For block $B_k$, three complementary rhythm statistics are computed --
(a) \textbf{absolute timing error:} mean $|\delta_r(o_i)|$,
(b) \textbf{signed bias:} mean $\delta_r(o_i)$, and
(c) \textbf{stability:} standard deviation of onset-level deviations
$\delta_r(o_i)$ within block $B_k$.

The aggregated rhythmic deviation is:
$
\delta_r^{(k)} =
\eta_1 \, \overline{|\delta_r|}^{(k)}
+
\eta_2 \, \mathrm{Std}^{(k)}
+
\eta_3 \, |\overline{\delta_r}^{(k)}|,
\quad \sum_i \eta_i = 1
$,
where within block $B_k$, $\overline{|\delta_r|}^{(k)}$ denotes the mean absolute
onset deviation,
$\mathrm{Std}^{(k)}$ or $\mathrm{Std}\left(\{\delta_r(o_i)\}_{o_i \in \mathbf{o}_k}\right)$ denotes the stability metric,
and
$\overline{\delta_r}^{(k)}$ denotes the signed mean deviation (bias).
The rhythm fidelity score is defined via bounded normalization:
$\mathcal{R}_k = 1 - \rho_r\left(\delta_r^{(k)}\right)$,
ensuring $\mathcal{R}_k \in [0,1]$.

\vspace{-0.4em}
\oursubsubsection{Block-Level Musical Consistency} \vspace{-0.3em}

The resulting $\mathcal{P}_k$ and $\mathcal{R}_k$ are fused to compute
the block-level musical score as described in Sec.~\ref{sec:block_level_fidelity_measures}.

\vspace{-0.4em}
\subsection{Overall Performance Evaluation}\label{sec:overall_performance_evaluation} \vspace{-0.2em}
\vspace{-0.2em}

\input{tables/compact_eval}

\input{tables/ablation_study}

\input{plots/transcription_perf_comparison}

\textbf{Quantitative Aggregation:\quad}
Block-level content and musical scores are aggregated using
Eq.~\ref{eq:overall_performance_score}.
For our experiments, we set $\gamma_{{}_\mathcal{C}} = 0.55$ and $\gamma_{{}_\mathcal{M}} = 0.45$,
placing slightly higher emphasis on lyrical fidelity.
These weights have been selected empirically based on validation-set
correlation with human expert ratings.

\noindent\textbf{Natural Language (NL) Feedback Generation:\quad}
In addition to the scalar score, we generate structured
natural-language feedback.
We provide an LLM with:
(a) ordered sequence $\{ \mathcal{C}_k \}_{k=1}^{K}$,
(b) ordered sequence $\{ \mathcal{M}_k \}_{k=1}^{K}$,
(c) ASR transcription $\hat{\ell}(t)$, and
(d) reference lyrics $\ell^{*}(t)$.
Block-wise score sequences preserve localized performance
variations (e.g., weaker chorus, stronger verse), enabling the production
of section-aware NL rather than relying
solely on the global aggregate.\looseness=-1

\input{tables/generalization_analysis}

\section{Experimental Results}
\vspace{-0.2em}

\textbf{Configuration:}
We conduct experiments on Linux workstation equipped with
2$\times$ NVIDIA Tesla V100-SXM2 GPUs (32\,GB each),
using GPU acceleration for ASR fine-tuning and inference.
We fine-tune \texttt{whisper-large-v3} using parameter-efficient
LoRA adapters ($r=16$, $\alpha=32$, dropout $=0.05$) applied to the
attention projection layers (\texttt{q\_proj}, \texttt{k\_proj},
\texttt{v\_proj}, \texttt{out\_proj}).
Audio inputs are limited to 12\,s with a maximum transcription length
of 256 tokens.
Training is performed for 10 epochs with a learning rate of
$10^{-4}$, batch-size 1 with 16-step gradient accumulation. The lyrics pipeline time-based ASR windows ($L=28$\,s, stride $=10$\,s) over proto-segments, discarding windows with $<25$ characters, block matching uses embedding/lexical/phonetic weights $(0.55,0.20,0.25)$ with threshold $0.72$. Musical analysis uses pYIN ($C2$--$C6$, frame $=2048$, hop $=256$) and onset detection ($\text{pre\_max}=3$, $\text{post\_max}=3$, $\delta=0.15$).

\noindent\textbf{Quantitative Validation:}
We sample a sequence of 120 vocal performances, scored by $\ge3$ human expert judges independently on a scale of 1-10 with the final score computed as the per-clip mean across judges and then assess these performances using {\ourmethod} to derive an overall score. Then, we rank them based on these two score sequences and derive two orderings. Table \ref{tab:compact_eval} shows how closely the ordering inferred by {\ourmethod} correlates with human expert ground truth. Evaluation on SingMOS-Pro is limited to models supporting lyrics/content evaluation ($\mathcal{C}$), as it lacks ground-truth for music score evaluation ($\mathcal{M}$).
On {\ourdata}, NL feedback via \texttt{gpt-oss-120b} \cite{agarwal2025gpt}
(Sec.~\ref{sec:overall_performance_evaluation})
yields a \texttt{all-MiniLM-L6-v2} \cite{wang2020minilm}
cosine similarity of 63.97
with expert comments.

\noindent\textbf{Component impact analysis:}
Table \ref{tab:ablation_study} presents key ablations. Table \ref{tab:lyrics_pitch_rhythm_ablation} showcases the impact of content and musical components on SQA. In an exemplary instance, where content $\mathcal{C}$, pitch $\mathcal{P}$, and rhythm $\mathcal{R}$, singularly emit scores 0.829, 0.490, and 0.491 respectively, the overall score is computed as: (a) 0.829 (for $\mathcal{C}$ only), (b) 0.490 (for $\mathcal{M}$ only; $55\mathpunct{:}45$ weightage), and (c) 0.677 (for $\mathcal{C} \land \mathcal{M}$; $55\mathpunct{:}25\mathpunct{:}20$ weightage). Aggregation of (c) bears the highest $\rho$ of 0.683. Table \ref{tab:lyrics_pipeline_ablation} further breaks down the components of the lyrics pipeline, proving that multi-signal block detection approach outperforms individual signals.
Fig. \ref{fig:transcription_perf_comparison} shows that the
singing transcription accuracy improves by $29.87\%$ due to MG-LoRA over the second best (averaged across {\ourdata}, SingMOS-Pro, and Jamendo ~\cite{durand-2023-contrastive}).
The base ASR $\rho$ of 0.518 improves to
0.583 (+$\mathcal{L}_{\text{ASR}}$),
0.597 (+$\mathcal{L}_{d}$),
0.616 (+$\mathcal{L}_{p}$),
0.622 (+$\mathcal{L}_{a}$),
0.626 (+$\mathcal{L}_{o}$),
showing 
a maximum benefit due to $\mathcal{L}_{p}$ after $\mathcal{L}_{\text{ASR}}$.

\noindent\textbf{Generalization of MG-LoRA:} Tables \ref{tab:genre_consistency}, \ref{tab:language_generalization} present evaluations across the top-5 {\ourdata} genres and 5 languages representing Whisper performance extremes.

\noindent\textbf{Qualitative Analysis:}
Table \ref{tab:qualitative_base_vs_finetuned} 
shows that MG-LoRA improves transcription in cases like sustained note prolongation, melisma, ornamentation (like \textit{gamakas}), portamento.
Our NL feedback is shown in Supplementary: 
\url{https://neelam472.github.io/MusicJudge/Supp.pdf} .

\input{tables/qualitative_base_vs_finetuned}

\vspace{-0.8em}
\section{Conclusion}
\vspace{-0.2em}

We introduce {\ourmethod} for automatic SQA,
providing a practical foundation for assistive training tools, synthetic music evaluation, and scalable judging support in music competitions.
On {\ourdata} and SingMOS-Pro, {\ourmethod} achieves Spearman $\rho = 0.683|0.483$, outperforming lyric-only and music-only baselines by $+31.9\%|+48.2\%$ and $+38.0\%|-$, respectively. Coupling linguistic and musical cues yields >9.1\% more reliable SQA than single-modality evaluation. Proposed multi-signal block detection further improves intra-song boundary localization ($\rho = 0.626$, $+2.96\%$ over the second best). Further, MG-LoRA improves lyric transcription robustness across genres ($20.1 \pm 7.52\%$ WER$\downarrow$) and languages ($27.7 \pm 10.87\%$ WER$\downarrow$).
Future work may explore diarization-aware multi-singer MG-LoRA modeling.\looseness=-1

\vspace{-0.8em}
\section{Generative AI Use Disclosure}
\vspace{-0.2em}
Research usage of \texttt{gpt-oss-120b} is for natural language feedback generation as described in Sec.~\ref{sec:overall_performance_evaluation} (examples presented in Supplementary). Other generative AI usage is strictly limited to permitted re-formatting of tables/plots.

\vspace{-0.8em}
\bibliographystyle{IEEEtran}
\bibliography{MusicJudge}

\end{document}

%% file: content/ours_names.tex
\providecommand{\ourmethod}{{\textsc{MusicJudge}}}
\providecommand{\ourdata}{{\textsc{SwaraLyrics}}}

\providecommand{\oursubsubsection}{\subsubsection}

%% file: content/authors.tex
\author[affiliation={1}]{Neelam}{Saini}
\author[affiliation={1}, orcid=0000-0003-1866-1408]{Sourav}{Ghosh}

\address{
    $^1$ Samsung R\&D Institute Bangalore, India
}

\email{neelam.saini@samsung.com, sourav.ghosh@samsung.com}

%% file: tables/compact_eval.tex
\input{content/ours_names}
\begin{table}[t]
\centering
\caption{\textbf{Agreement and error metrics for singing evaluation.} 
{\scriptsize
$\rho$ = Spearman \cite{spearman1961proof}, $\tau$ = Kendall \cite{kendall1938new}. 
Lower MSE \cite{fisher1922mathematical}, MAE \cite{fisher1920mathematical}, and MedAE \cite{hampel1974influence} are better ($\downarrow$).
}\vspace{-1.1em}}
\label{tab:compact_eval}

\resizebox{\columnwidth}{!}{
\begin{tabular}{l|cc|ccccc|ccccc}
\toprule

  \multirow{2}{*}{\textbf{Method}}
& \multirow{2}{*}{$\mathcal{C}$}
& \multirow{2}{*}{$\mathcal{M}$}
& \multicolumn{5}{c|}{\textbf{\ourdata}} & \multicolumn{5}{c}{\textbf{SingMOS-Pro} \cite{tang2025singmos}} \\\cline{4-13}\noalign{\vspace{0.5ex}}

& &
& $\mathbf{\rho} \uparrow$ & $\mathbf{\tau} \uparrow$ & \textbf{MSE} $\downarrow$ & \textbf{MAE} $\downarrow$ & \textbf{MedAE} $\downarrow$
& $\mathbf{\rho} \uparrow$ & $\mathbf{\tau} \uparrow$ & \textbf{MSE} $\downarrow$ & \textbf{MAE} $\downarrow$ & \textbf{MedAE} $\downarrow$ \\

\midrule

SingMOS \cite{tang2024singmos}  & \ding{51} & \ding{55}
& - & - & - & - & -
& $0.091$ & $0.062$ & $0.56212$ & $0.60370$ & $0.45039$\\

UTMOS \cite{saeki22c_interspeech}  & \ding{51} & \ding{55}
& - & - & - & - & -
& $0.120$ & $0.076$ & $0.24000$ & $0.39200$ & $0.29400$\\

DNSMOS \cite{reddy2021dnsmos}  & \ding{51} & \ding{55}
& - & - & - & - & -
& $0.201$ & $0.137$ & $0.07560$ & $0.22000$ & $0.16500$\\

Whisper \cite{radford2023robust}  & \ding{51} & \ding{55}
& $0.518$& $0.350$& $0.00960$& $0.08010$& $0.06250$
& $0.326$ & $0.241$ & $0.06829$ & $0.20020$ & $0.16600$\\

\rowcolor{gray!20}
+ {\hfill MG-LoRA}   & \ding{51} & \ding{55}
& $\mathbf{0.626}$& $\mathbf{0.459}$& $\mathbf{0.00685}$& $\mathbf{0.06073}$& $\mathbf{0.04250}$
& $\mathbf{0.483}$ & $\mathbf{0.379}$ & $\mathbf{0.04275}$ & $\mathbf{0.15129}$ & $\mathbf{0.10799}$\\

\midrule

SWIPE \cite{camacho2008sawtooth}  & \ding{55} & \ding{51}
& $0.455$& $0.320$& $0.00910$& $0.07600$& $0.06500$
& $\times$ & $\times$ & $\times$ & $\times$ & $\times$\\

CREPE \cite{kim2018crepe}  & \ding{55} & \ding{51}
& $0.482$& $0.345$& $0.00870$& $0.07400$& $0.06300$
& $\times$ & $\times$ & $\times$ & $\times$ & $\times$\\

pYIN \cite{6853678} & \ding{55} & \ding{51}
& $\mathbf{0.495}$& $\mathbf{0.354}$& $\mathbf{0.00836}$& $\mathbf{0.06673}$& $\mathbf{0.03600}$
& $\times$ & $\times$ & $\times$ & $\times$ & $\times$\\

\midrule

\rowcolor{gray!20}
{\ourmethod}   & \ding{51} & \ding{51}
& $\mathbf{0.683}$& $\mathbf{0.499}$& $\mathbf{0.00564}$& $\mathbf{0.05514}$& $\mathbf{0.03633}$
& $\mathbf{0.483}$ & $\mathbf{0.379}$ & $\mathbf{0.04275}$ & $\mathbf{0.15129}$ & $\mathbf{0.10799}$\\

\bottomrule
\end{tabular}
}
\end{table}

%% file: tables/ablation_study.tex
\input{content/ours_names}
\begin{table}[t]
    \centering
    \caption{{\ourmethod} Ablation Study on {\ourdata}\vspace{-2ex}}
    \label{tab:ablation_study}
    \begin{subtable}[t]{0.52\columnwidth}
        \centering
        \caption{Content vs. Pitch-Rhythm\vspace{-2ex}}
        \label{tab:lyrics_pitch_rhythm_ablation}
        \resizebox{\columnwidth}{!}{%
        \begin{tabular}{l|cc}
            \toprule
            \textbf{Configuration} &
            \thead{\boldmath$\rho \uparrow$} &
            \thead{\textbf{MSE} $\downarrow$} \\
            \midrule
            Musical Score $\mathcal{M}$ only & $0.495$ & $0.00836$ \\
            Content Score $\mathcal{C}$ only & $0.626$ & $0.00685$ \\
            Both ($\mathcal{C} \land \mathcal{M}$) & $\mathbf{0.683}$ & $\mathbf{0.00564}$ \\
            \bottomrule
        \end{tabular}%
        }
    \end{subtable}
    \hfill
    \begin{subtable}[t]{0.46\columnwidth}
        \centering
        \caption{Multi-signal (Sec. \ref{sec:multi_signal_block_detection})\vspace{-1.8ex}}
        \label{tab:lyrics_pipeline_ablation}
        \resizebox{0.95\columnwidth}{!}{%
        \begin{tabular}{l|cc|c}
            \toprule
            \textbf{Variant} & 
            \boldmath$\alpha_{\text{embed}}$ & 
            \boldmath$\alpha_{\text{fuzzy}}$ & 
            \thead{\boldmath$\rho \uparrow$} \\
            \midrule
            \texttt{NO\_EMBED} & $0.00$ & $0.5$ & $0.495$ \\
            \texttt{NO\_PHONETIC} & $0.70$ & $0.3$ & $0.560$ \\
            \texttt{NO\_FUZZY} & $0.70$ & $0.0$ & $0.608$ \\
            \texttt{FULL\_ALL} & $0.55$ & $0.2$ & $\mathbf{0.626}$ \\
            \bottomrule
        \end{tabular}%
        }
    \end{subtable}
    \vspace{-2em}
\end{table}

%% file: plots/transcription_perf_comparison.tex
\input{content/ours_names}
\begin{figure}[t]
\centering
\begin{tikzpicture}
\begin{axis}[
    width=\columnwidth,
    height=2.85cm,
    ybar,
    bar width=5pt,
    ymin=0,
    ylabel={WER \cite{hunt1990figures}},
    symbolic x coords={
        A,
        B,
        C,
        D,
        E (Ours)
    },
    xtick=data,
    x tick label style={rotate=0,font=\scriptsize},
    enlarge x limits=0.18,
    legend style={
        font=\tiny,
        at={(0.5,1.25)},
        anchor=south,
        legend columns=3
    },
    font=\tiny,
]

\addplot coordinates {
    (A,1.2250)
    (B,1.1500)
    (C,0.8500)
    (D,0.7400)
    (E (Ours),0.6100)
};

\addplot coordinates {
    (A,1.0674)
    (B,1.0559)
    (C,0.6620)
    (D,0.4052)
    (E (Ours),0.2218)
};

\addplot coordinates {
    (A,1.2613)
    (B,1.2162)
    (C,0.7748)
    (D,0.7477)
    (E (Ours),0.5474)
};

\legend{SingMOS WER, Jamendo WER, {\ourdata} WER}

\end{axis}

\begin{axis}[
    width=\columnwidth,
    height=2.85cm,
    ymin=0,
    axis y line*=right,
    axis x line=none,
    ylabel={CER \cite{morris04_interspeech}},
    symbolic x coords={
        A,
        B,
        C,
        D,
        E (Ours)
    },
    xtick=\empty,
    legend style={
        font=\tiny,
        at={(0.5,1)},
        anchor=south,
        legend columns=3
    },
    font=\tiny,
]

\addplot+[mark=*] coordinates {
    (A,0.3405)
    (B,0.2830)
    (C,0.2040)
    (D,0.1990)
    (E (Ours),0.1062)
};

\addplot+[mark=square*] coordinates {
    (A,0.8952)
    (B,0.9113)
    (C,0.4421)
    (D,0.2627)
    (E (Ours),0.2234)
};

\addplot+[mark=triangle*] coordinates {
    (A,1.1030)
    (B,1.0730)
    (C,0.5056)
    (D,0.4850)
    (E (Ours),0.4382)
};

\legend{SingMOS CER, Jamendo CER, {\ourdata} CER}

\end{axis}
\end{tikzpicture}
\vspace{-2.2em}
\caption{
\textbf{Singing ASR Performance.} Lower is better.
\textnormal{\scriptsize
A: \texttt{hubert-large-ls960-ft} \cite{hsu2021hubert},
B: \texttt{wav2vec2-large-960h-lv60} \cite{baevski2020wav2vec},
C: \texttt{whisper-medium},
D: \texttt{whisper-large-v3},
E: D + MG-LoRA.
}
\vspace{-2em}
}
\label{fig:transcription_perf_comparison}
\end{figure}
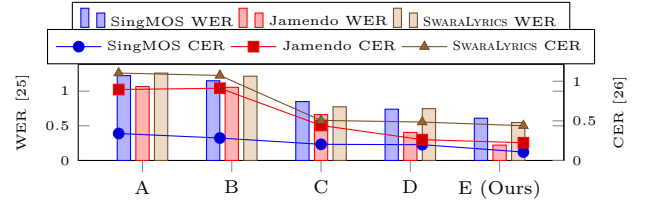

%% file: tables/generalization_analysis.tex
\input{content/ours_names}
\begin{table}[h]
    \vspace{-2ex}
    \centering
    \caption{MG-LoRA transcription robustness}
    \vspace{-2.5ex}
    \label{tab:generalization_analysis}
    \begin{subtable}[t]{0.46\columnwidth}
        \centering
        \caption{across singing genres}
        \vspace{-1.8ex}
        \label{tab:genre_consistency}
        \resizebox{\columnwidth}{!}{
        \begin{tabular}{l|cc|cc}
            \toprule
            \multirow{2}{*}{\textbf{Genre}} & \multicolumn{2}{c|}{\textbf{Base}} & \multicolumn{2}{c}{\textbf{MG-LoRA}} \\\cline{2-5}
            & \textbf{WER} & \textbf{CER} & \textbf{WER} & \textbf{CER}\\
            \midrule
            Classical & $0.800$ & $0.671$ & $\mathbf{0.689}$ & $\mathbf{0.563}$ \\
            Folk      & $0.742$ & $0.624$ & $\mathbf{0.497}$ & $\mathbf{0.405}$ \\
            Ghazal    & $0.682$ & $0.592$ & $\mathbf{0.571}$ & $\mathbf{0.482}$ \\
            Bhajan    & $0.642$ & $0.534$ & $\mathbf{0.529}$ & $\mathbf{0.421}$ \\
            Pop       & $0.562$ & $0.423$ & $\mathbf{0.451}$ & $\mathbf{0.319}$ \\
            \bottomrule
        \end{tabular}}
        \label{tab:genre_asr_results_final}
    \end{subtable}
    \hfill
    \begin{subtable}[t]{0.50\columnwidth}
        \centering
        \caption{across languages}
        \vspace{-1.8ex}
        \label{tab:language_generalization}
        \resizebox{\columnwidth}{!}{%
        \begin{tabular}{l|cc|cc}
            \toprule
            \multirow{2}{*}{\textbf{\small Language}} & \multicolumn{2}{c|}{\textbf{Base}} & \multicolumn{2}{c}{\textbf{MG-LoRA}} \\\cline{2-5}
            & \textbf{WER} & \textbf{CER} & \textbf{WER} & \textbf{CER}\\
            \midrule
            English    & $0.4052$ & $0.2627$ & $\mathbf{0.2218}$ & $\mathbf{0.2234}$ \\
            Mandarin   & $0.7400$ & $0.1990$ & $\mathbf{0.6100}$ & $\mathbf{0.1062}$ \\
            Hindi      & $0.7477$ & $0.4850$ & $\mathbf{0.5474}$ & $\mathbf{0.4382}$ \\
            Punjabi    & $0.9431$ & $0.6347$ & $\mathbf{0.6705}$ & $\mathbf{0.3854}$ \\
            Bengali    & $0.9375$ & $0.5153$ & $\mathbf{0.7500}$ & $\mathbf{0.4365}$ \\
            \bottomrule
        \end{tabular}
        }
    \end{subtable}
    \vspace{-2.2em}
\end{table}

%% file: tables/qualitative_base_vs_finetuned.tex
\input{content/ours_names}
\begin{table}[ht]
    \centering
    \vspace{-1em}
    \caption{Exemplary instances of lyrics transcription}
    \vspace{-1em}
    \label{tab:qualitative_base_vs_finetuned}
    \resizebox{\columnwidth}{!}{%
    \begin{tabular}{lll}
        \toprule
        \textbf{Ground Truth Lyrics} & \textbf{Whisper (Base)} & \textbf{Whisper + MG-LoRA (Ours)} \\
        \midrule

        \begin{tabular}[c]{@{}l@{}}
        {\devanagarifont उलझन मेरी सुलझा दे, चाहूँ मैं या ना} \\
        (Uljhan meri suljha de, \\
        chahoon main aana)
        \end{tabular}
        &
        \begin{tabular}[c]{@{}l@{}}
        {\devanagarifont \textcolor{red}{\uline{उलज़म}} मेरी \textcolor{red}{\uline{सुलजा}} दे \textcolor{red}{\uline{चाहो}} मैं \textcolor{red}{\uline{आना}}} \\
        (\textcolor{red}{\uline{Uljham}} meri \textcolor{red}{\uline{sulja}} de \\
        \textcolor{red}{\uline{chaho}} main \textcolor{red}{\uline{aana}})
        \end{tabular}
        &
        \begin{tabular}[c]{@{}l@{}}
        {\devanagarifont \textcolor{ForestGreen}{\uline{उलझन}} मेरी \textcolor{red}{\uline{सुलजा}} दे, \textcolor{ForestGreen}{\uline{चाहूँ}} मैं या ना} \\
        (\textcolor{ForestGreen}{\uline{Uljhan}} meri \textcolor{red}{\uline{sulja}} de, \\
        \textcolor{ForestGreen}{\uline{chahoon}} main \textcolor{ForestGreen}{\uline{ya na}})
        \end{tabular}
        \\

        \addlinespace\midrule

        \begin{tabular}[c]{@{}l@{}}
        {\devanagarifont मेरा कोई एहसास है जैसे} \\
        (Mera koi ehsaas hai jaise)
        \end{tabular}
        &
        \begin{tabular}[c]{@{}l@{}}
        {\devanagarifont मेरा कोई \textcolor{red}{\uline{अहसास}} है \textcolor{red}{\uline{जेसे}}} \\
        (Mera koi \textcolor{red}{\uline{ahsaas}} hai \textcolor{red}{\uline{jese}})
        \end{tabular}
        &
        \begin{tabular}[c]{@{}l@{}}
        {\devanagarifont मेरा कोई \textcolor{ForestGreen}{\uline{एहसास}} है \textcolor{ForestGreen}{\uline{जैसे}}} \\
        (Mera koi \textcolor{ForestGreen}{\uline{ehsaas}} hai \textcolor{ForestGreen}{\uline{jaise}})
        \end{tabular}
        \\

        \addlinespace\midrule

        \begin{tabular}[c]{@{}l@{}}
        {\devanagarifont सीने से तुम मेरे आ के लग जाओ ना} \\
        {\devanagarifont डरते हो क्यूँ?} \\ 
        {\devanagarifont ज़रा पास तो आओ ना} \\
        (Seene se tum mere aa ke lag \\
        jao na darte ho kyun? \\
        Zara paas to aao na)
        \end{tabular}
        &
        \begin{tabular}[c]{@{}l@{}}
        {\devanagarifont सीने से तुम मेरे \textcolor{red}{\uline{आखे}} लग जाओ ना} \\
        {\devanagarifont \textcolor{red}{\uline{दर्दे}} हो \textcolor{red}{\uline{क्यों}}}\\
        {\devanagarifont \textcolor{red}{\uline{जरा}} पास तो आओ ना} \\
        (Seene se tum mere \textcolor{red}{\uline{aakhe}} lag \\
        jao na \textcolor{red}{\uline{darde}} ho \textcolor{red}{\uline{kyon}} \\
        \textcolor{red}{\uline{jara}} paas to aao na)
        \end{tabular}
        &
        \begin{tabular}[c]{@{}l@{}}
        \texttt{\devanagarifont सीने से तुम मेरे \textcolor{ForestGreen}{\uline{आ के}} लग जाओ ना,} \\
        {\devanagarifont \textcolor{red}{\uline{दरते}} हो \textcolor{ForestGreen}{\uline{क्यूँ}},} \\
        {\devanagarifont \textcolor{ForestGreen}{\uline{ज़रा}} पास तो आओ ना} \\
        (Seene se tum mere \textcolor{ForestGreen}{\uline{aa ke}} lag \\
        jao na, \textcolor{ForestGreen}{\uline{darte}} ho \textcolor{ForestGreen}{\uline{kyun}}, \\
        \textcolor{ForestGreen}{\uline{zara}} paas to aao na)
        \end{tabular}
        \\

        \bottomrule
    \end{tabular}%
    
    }
    \vspace{-1.4em}
\end{table}